\documentclass[aps,prc,showpacs,floatfix,twocolumn]{revtex4}

\usepackage{amsmath}
\ifx\pdftexversion\undefined
  \usepackage[dvips]{graphicx,color}
\else
  \usepackage[pdftex]{graphicx,color}
\fi

\def\pT{\mbox{$p_T\!$ }}
\def\v2{\mbox{$v_2$}}

\def\eq#1{{Eq.~(\ref{#1})}}
\def\fig#1{{Fig.~\ref{#1}}}

\begin{document}

%
\hyphenation{author another created financial paper re-commend-ed Post-Script}

\title{ Decomposition of Harmonic and Jet Contributions to Particle-pair Correlations at Ultra-relativistic Energies }
\author{ N. N. Ajitanand, J. M. Alexander, P. Chung, W. G. Holzmann, 
M. Issah, Roy A. Lacey, A. Shevel and A. Taranenko 
}
\affiliation{ Department of Chemistry, State University of New York at Stony Brook, Stony 
Brook, NY 11794  
}
\author{P.~Danielewicz }
\affiliation{Michigan State University, East Lansing MI 48824-1321 }
%
%
\date{\today}

\begin{abstract}
Methodology is presented for analysis of two-particle azimuthal angle correlation functions
obtained in collisions at ultra-relativistic energies. We show that harmonic and 
di-jet contributions to these correlation functions can be reliably decomposed by two 
techniques to give an accurate measurement of the jet-pair distribution. 
Results from detailed Monte Carlo simulations are used to demonstrate the efficacy 
of these techniques in the study of possible modifications to jet topologies in 
heavy ion reactions. 
\end{abstract}

\pacs{PACS 25.75.Ld}
\maketitle



A primary objective for studying ultra-relativistic collisions between 
complex nuclei is the creation and characterization of a new phase of hot and dense 
nuclear matter, where quarks and gluons are no longer confined to the interior of 
single hadrons. The existence of this Quark-Gluon Plasma (QGP), was conjectured based 
on the observation of asymptotic freedom~\cite{Gross:1973id,Gross:1973ju} and is 
now very strongly indicated by lattice QCD 
calculations~\cite{Redlich:2004gp,Fodor:2001pe,Fodor:2004nz}.
The energy loss of hard scattered partons in this de-confined medium is predicted 
to be larger than that for cold hadronic matter \cite{Bjorken:1982tu,Gyulassy:2003mc}. 
Such an energy loss can result in an apparent softening of the fragmentation function 
for jets, leading to suppression in the observed yield of high \pT hadrons and to 
modifications of jet topologies. Thus, a careful study of jet properties and yields in heavy ion 
collisions provides a sensitive probe for possible 
QGP formation \cite{Gyulassy:2003mc,Wang:2004hy} and possibly for QGP properties. 

It is known 
that characteristic back-to-back jets of hadrons are copiously emitted in 
sufficiently energetic collisions. Studies of these jets provide information on  
related fragmentation functions in a ``cold'' nuclear 
medium\cite{Stewart:1990wa,Corcoran:1990vq,Naples:1994uz}.
The more recent program of 
study at the Relativistic Heavy Ion Collider (RHIC) has involved p-p, d-Au, 
and Au-Au collisions. It is conjectured that the first produces jets in free space, the second 
produces jets that could be influenced by relatively modest initial state 
effects, and the third produces jets that could be strongly influenced by 
final-state effects such as those resulting from a reaction zone of very 
high energy-density nuclear matter. Hadron pair correlations provide a 
primary tool for the study of the yield and topology of these jets. 
Therefore, much excitement revolves around the observation, 
comparison, and characterization of such correlations in these three 
situations \cite{Ajitanand:2002qd,Jacobs:2003bx,Adler:2002tq,Rak:2004gk,Adler:2004zd}. 

Pair correlation studies of d-Au and Au-Au collisions have to deal with a relatively 
large ``underlying event''. For d-Au reactions, this underlying event is 
uncorrelated with the reaction plane (RP). By contrast, for Au-Au reactions 
there are rather strong correlations between the reaction plane 
and the azimuthal emission angle of 
hadrons \cite{Ajitanand:2002qd,Adler:2002ct,Adler:2003kt,Ackermann:2000tr}. 
This ``elliptic flow'' leads to particle-plane-particle or ``harmonic'' pair 
correlations that must be taken into account as one tries to identify and 
characterize jet pair correlations. Flow correlations are presumed 
to result primarily from pressure gradients developed in the initial anisotropic (or almond 
shaped) overlap region of high energy-density nuclear 
matter\cite{Ollitrault:1992bk,Danielewicz:2002pu,Teaney:2001av,Kolb:2001qz,Hirano:2004rs}. 
The interactions within this overlap region may also alter jet properties, and one has 
to allow for this possibility in any analysis. 

	Recent experimental studies of jet properties have focused on 
the decomposition of $\Delta\phi$ 
distributions \cite{Adler:2002tq,Bielcikova:2004jf,Adams:2004wz,Adler:2004zd}.	
Here, we present two methods which exploit the use of correlation functions 
to separate jet driven pair correlations from harmonic correlations. 
The use of correlation functions circumvents the need for full 
azimuthal detector acceptance and can serve to minimize many 
important systematic uncertainties (acceptance, efficiency, etc) which 
could influence the accuracy of extracted jet properties.
A unique capability of one of the methods presented is the direct study of  
modifications to jet properties via the extinction of harmonic correlations.
The methods are illustrated and tested via detailed Monte Carlo simulations. 
Applications to the analysis of RHIC data are presented elsewhere. 


{\bf Azimuthal angle correlation functions:}
To emphasize jet correlations, one often selects events with at least one 
high-$p_T$ particle. This ``trigger particle'' is then paired 
with other ``associated'' particles to 
obtain azimuthal angle pair correlations. Hereafter, we refer to the trigger 
and associated particles as A and B respectively. The AB 
correlations are termed ``fixed correlations'' if A and B are from the 
same \pT bin or ``assorted correlations'' if A and B are from separate 
\pT bins. AB pairs are used to construct a correlation function $C^{AB}(\Delta\phi$) in 
azimuthal angle difference $\Delta \phi = \left(\phi _{A }-\phi _{B}\right)$; this function is 
constructed by dividing an area normalized (foreground) distribution of real particle pairs by an 
area normalized (background) distribution of mixed pairs obtained by pairing 
particles from different events. 

The foreground distribution F is defined in terms of the number of 
coincident (or real) pairs N$^{AB}$ per event;
\begin{equation}
F(\Delta \phi ) = N^{AB} (\Delta \phi ) / \Sigma  N^{AB} 
(\Delta \phi ), 
\label{eq1}
\end{equation}
where the sum is taken over the bins in $\Delta \phi$.

This gives the probability distribution for detecting an AB pair in a bin of 
$\Delta \phi$. The background distribution is defined in the same way, 
but by selecting A and B particles from separate events to form pairs;
\begin{equation}
B(\Delta \phi ) = [N^{AB} (\Delta \phi ) / \Sigma  N^{AB} 
(\Delta \phi )]_{mix}. 
\label{eq2}
\end{equation}
This gives the detection efficiency for an AB pair in the bin of $\Delta 
\phi $. The ratio of these two distributions defines the correlation 
function
\begin{equation}
C(\Delta \phi ) = F(\Delta \phi )/B(\Delta \phi ). 
\label{eq3}
\end{equation}
This correlation function gives the efficiency corrected probability distribution for real 
coincident pairs.

\begin{figure}
 \includegraphics[width=1.0\linewidth]{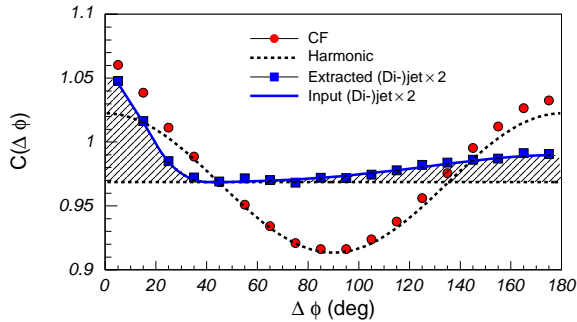}  
 \caption{\label{fig1}
	Correlation function from simulated data (filled circles). 
The dotted curve shows the input harmonic component. The solid curve and 
full squares show input and output jet pair distributions referenced to an offset.
}
 \end{figure}

The filled circles in \fig{fig1} show a typical correlation function generated from 
simulated data tuned to recent experimental data from 
RHIC \cite{Adler:2002ct,Adcox:2002ms}. There is a relatively narrow peak 
at 0$^{o}$, a broader peak at 180$^{o}$, and a minimum at $\sim $ 80$^{o}$. 
Our task is to devise an analysis method that allows one to retrieve the 
input jet-pair distribution used to generate the correlation function. 

{\bf Two source model:}
For Au-Au reactions at RHIC it has been shown that a dominant aspect of 
hadron pair correlations is collective flow of second order, i.e. elliptic 
flow \cite{Adler:2002ct,Adcox:2002ms}. 
This gives a harmonic distribution for particle type A (or B) 
with respect to the angle of the reaction plane $\psi _{R}$.
\begin{equation}
N^{A} (\phi _{A} - \psi _{R}) \propto [1 + 2 v_{2}^{A} 
cos 2 (\phi _{A} - \psi _{R})]
\label{eq4}
\end{equation}
and a corresponding harmonic correlation function between AB pairs \cite{Wang:1991qh,Lacey:1993cf}
\begin{equation}
C_{H}^{AB} (\Delta \phi )_{ } = [1 + 2 v_{2} 
cos 2(\Delta \phi _{ }) ]; \qquad v_{2}=(v_{2}^{A}\times v_{2}^{B}).
\label{eq5}
\end{equation}
In addition to these harmonic correlations one expects characteristic 
contributions from jets. To simulate both source contributions we include both 
harmonic and di-jet particle emission patterns into a Monte Carlo code. 
The simulations use exponential \pT distributions and Poisson 
sampling for the number of jets per event, the number of particles per jet 
and the number of harmonically flowing particles per event. The average 
number of near and away-side jet particles were set to be equal, and the 
near and away-side jets were generated with an effective $j_T $ and $k_T $ 
respectively. Here, $j_T $ reflects the average transverse momentum projection 
of hadrons perpendicular to the nearside jet axis 
$\left( {j_T =p_T^{hadron} \sin \left( {\phi _{jet-had} } \right)} \right)$ and $k_T$ 
reflects the acoplanarity of the near- and away-side jets,  
$\left( {k_T = (p_T^{jet}/\sqrt(2)) \sin \left( {\Delta \phi _{jet-jet}} \right)} \right)$. 
Input parameters were adjusted to yield correlation functions in rough agreement 
with the measured ones. An absorption factor was introduced to study 
the effects of possible suppression of the away-side jet (J$_{a})$. The axis 
of the nearside jet (J$_{n})$ was also made to correlate with varying strength to 
the RP via a harmonic function (as in. \eq{eq4}) or left uncorrelated. 

It can be shown~\cite{stankus_Qm} that the pair correlations from such flow and jet 
sources are given by 
\begin{equation}
C^{AB} (\Delta \phi ) = a_{o} [C^{AB}_H(\Delta \phi ) + J (\Delta \phi )], 
\label{eq6}
\end{equation}
where $C^{AB}_H(\Delta \phi )$ is a harmonic function of effective amplitude 
v$_{2}$, \eq{eq5}, and J($\Delta \phi )$ is a di-jet function. 
No explicit or implicit assumption is made for the functional form of 
of $J$. In \fig{fig1} the dot-dashed and solid curves represent 
the input harmonic and jet functions respectively.

%
By rearrangement of \eq{eq6} one obtains 
\begin{equation}
J(\Delta \phi ) = [C^{AB}(\Delta \phi ) - a_{o}C^{AB}_{H} (\Delta \phi )] / a_{o}, 
\label{eq7}
\end{equation}
Knowledge of a$_{o}$ is required to evaluate $J$. Next we discuss an 
ansatz for assigning the value of a$_{o}$.

From measurements of the di-jet functions in d+Au reactions, one finds that 
they are very similar to the ones obtained in p+p reactions and can be reasonably described by a 
double Gaussian function with relatively narrow near- and away-side widths 
($\sigma _{n}$ and $\sigma _{a }$ respectively)~\cite{Rak:2004gk,unknown:2005ph}. For similar 
\pT selection, Au+Au reactions are found to give near-side jet widths 
which are similar to those obtained in d+Au. By contrast, the away-side jet 
seems to broaden and to diminish in amplitude as the centrality is increased 
\cite{Rak:2004gk}. Guided by these observations it is reasonable to assume, as a starting 
point, that the di-jet function in Au+Au reactions has negligible intensity 
at the minimum $\Delta \phi _{min}$, in the jet function (cf. \fig{fig1}), i.e.
\begin{equation}
a_0C_{H}^{AB}(\Delta{\phi}_{min}) = C^{AB}(\Delta \phi_{min}),
\label{eq8}
\end{equation}
which can be solved to obtain $a_0$.
Thus the condition, zero yield at minimum (ZYAM), serves to fix the normalization 
constant $a_{o}$. 

	From the two source model described above, one obtains a single particle distribution, 
with respect to the reaction plane $\psi_R$ which can be fit with a harmonic. 
Its amplitude $v_2$, reflects an average over the particles from both sources. 
For example, if one simulates a flow source (90\%) with $v_2 = 0.2$ and an 
uncorrelated jet source (10\%), then the observed value of $v_2$ will 
be $0.9\times 0.2 + 0.1\times 0$ or 0.18. A finite resolution in the 
determination of $\psi _{R}$ requires a correction factor in the evaluation of $v_2$ 
as described in Refs.~\cite{dan85,olli98,Borghini:2001vi,Poskanzer:1998yz}. 
Knowledge of $v_2$ determines $C^{AB}_H$ in Eq.~\ref{eq6}. The determination of 
the effective $v_2$ utilizing the reaction plane as discussed above, requires that 
the reaction plane is itself determined by a procedure free of non-flow effects.
One such example is a measurement which demands a large (pseudo)rapidity gap ($\sim 3 - 4$ units) 
between the reaction plane and the particles correlated with it~\cite{Adcox:2002ms,Adler:2004cj}. 
It is expected that the $v_2$ values so obtained are much less affected by jet  
contributions~\cite{Adcox:2002ms,Adler:2004cj}. 
Earlier work has shown that $v_2$ varies with
both centrality and $p_T$. Therefore, to avoid biases, one must measure $v_2^A$
and $v_2^B$ for exactly the same data set to be analyzed for its jet function. To
accomplish this, one also uses the reaction planes and their resolution corrections for 
exactly the same events of interest and determines $v_2^A$ and $v_2^B$ from Eq.~\ref{eq4} for 
the particle of interest.
From Eq.~\ref{eq5} one then obtains the effective amplitude $v_2$.

	With the ZYAM condition and the measured values of $v_2$, one can decompose the 
observed correlation function into jet and harmonic components. This is illustrated in
Fig.~\ref{fig1}. The solid circles represent the simulated correlation data. The dot-dashed
curve represents the effective harmonic component about $a_0$. 
The filled triangles show the difference with an $a_0$ offset, which can be compared 
to the input jet function (also with an $a_0$ offset) represented by the solid line. 
Note that the values of $v_2^A$ and $v_2^B$ are obtained from fits to the azimuthal
distributions with respect to the reaction plane.

\begin{figure}
 \includegraphics[width=1.0\linewidth]{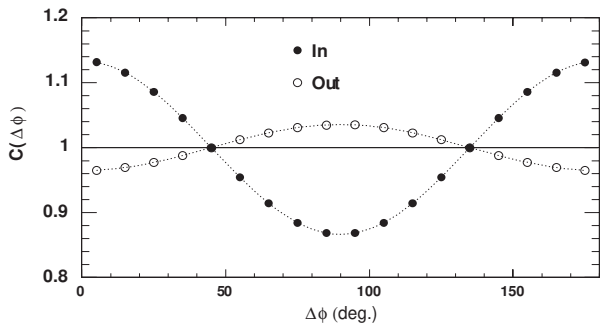}  
 \caption{\label{fig2}
	Simulated data, closed and open circles, for inplane C$^{in}$ and out of plane C$^{out}$ 
correlation functions as described in the text with $\Delta \phi _{c}$ = 
45$^{o}$ and $\Delta \psi _{R}$ = 40$^{o}$. Only harmonic correlations 
are present as shown by the dotted curve with amplitude v$_{2}^A$ = v$_{2}^B$.
}
 \end{figure}

\begin{figure}
 \includegraphics[width=1.0\linewidth]{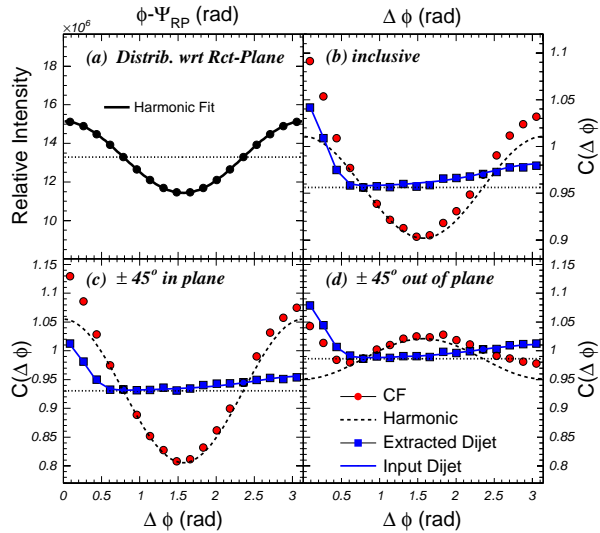}  
 \caption{\label{fig3}
	(a) Simulated data points and smooth curve for the correlation with respect to
the reaction plane. (b) Simulated data (filled circles) for an inclusive 
correlation function including harmonic and jet correlations . 
The dotted curve indicates the harmonic term. The solid curve and full squares 
show input and output jet distributions referenced to a$_{o}$. (c) Same as (b) 
but for an in-plane correlation function Eq.(9). (d) Same as (b) but for an 
out-of-plane correlation function Eq.(10). 
}
 \end{figure}

	The method can be extended to obtain information on the correlation of the jet function
with respect to the reaction plane~\cite{Bielcikova:2003ku,Borghini:2004ra}.
Two new correlation functions are constructed as 
follows: (1) the trigger particle $A$ is constrained to have its $\phi $ angle 
within a certain cut-angle $\Delta \phi _{c}$, about the reaction plane, and the associated 
particle $B$ is unconstrained. 
The mixed-event background $B(\Delta \phi )$ is constructed as before in \eq{eq3},
\begin{equation}
C^{in}(\Delta \phi ) = F^{in}(\Delta \phi )/B(\Delta \phi).
\label{eq9}
\end{equation}
	(2) The trigger particle is constrained to have its $\phi$ angle within 
$\Delta \phi _{c }$ of the normal to the reaction plane
\begin{equation}
C^{out} (\Delta \phi ) = F^{out} (\Delta \phi ) / B(\Delta\phi ).
\label{eq10}
\end{equation}
Figure \ref{fig2} shows the effect of these constraints on simulated data for pure 
flow with no di-jet contribution. Despite a sizable dispersion of the reaction plane 
($\Delta \psi _{R}$ = 40$^{o}$), Fig.~\ref{fig2} shows that the effects of these 
constraints are strong and are in opposing directions for C$^{in}(\Delta \phi )$ versus 
C$^{out}(\Delta \phi )$. The smooth curves in the figure show results 
derived from the analytical formulas developed in Ref.~\cite{Bielcikova:2003ku} for the 
relationship between (v$_{2}^A)^{in}$ and (v$_{2}^A)^{out}$ for given values of 
v$_{2}^A$, the reaction plane resolution $\Delta \psi _{R}$, and the 
angular selection $\Delta \phi_{c}$ used to constrain the direction of 
the trigger particle~\cite{Bielcikova:2003ku}.
\begin{widetext}
\begin{eqnarray}
(v_2^A)^{out} =\left( {\frac{2v_2^A \left( {\Delta \phi _c } \right)-\sin \left( 
{2\Delta \phi _c } \right)\left\langle {\cos \left( {2\Delta \Psi _R } 
\right)} \right\rangle +\frac{v_2^A }{2}\sin \left( {4\Delta \phi _c } 
\right)\left\langle {\cos \left( {4\Delta \Psi _R } \right)} \right\rangle 
}{2\left( {\Delta \phi _c } \right)-2v_2^A \sin \left( {2\Delta \phi _c } 
\right)\left\langle {\cos \left( {2\Delta \Psi _R } \right)} \right\rangle 
}} \right)
\label{eq11}
\\
(v_2^A)^{in} =\left( {\frac{2v_2^A \left( {\Delta \phi _c } \right)+\sin \left( 
{2\Delta \phi _c } \right)\left\langle {\cos \left( {2\Delta \Psi _R } 
\right)} \right\rangle +\frac{v_2^A }{2}\sin \left( {4\Delta \phi _c } 
\right)\left\langle {\cos \left( {4\Delta \Psi _R } \right)} \right\rangle 
}{2\left( {\Delta \phi _c } \right)+2v_2^A \sin \left( {2\Delta \phi _c } 
\right)\left\langle {\cos \left( {2\Delta \Psi _R } \right)} \right\rangle 
}} \right)
\label{eq12}
\end{eqnarray}
\end{widetext}
The harmonic amplitude $v_2$ in Eq.~\ref{eq5} is replaced by $(v_2^A)^{in}(v_2^B)$ and by 
$(v_2^A)^{out}(v_2^B)$. In the presence of jets, one can also use the ZYAM condition 
to decompose in-plane or out-of-plane correlations.
This is illustrated in Fig.~\ref{fig3} where results are generated that include 
harmonic emissions of 92\% of the particles and di-jet emissions of 8\%. Panel (a)
shows simulated data points compared to a fit by Eqn.~\ref{eq4} for the particle distribution
with respect to the reaction plane. This fit gives the effective value of $v_2$ to be 
used in the analysis of the two-particle correlation functions. Panels (b), (c), and (d) show 
simulated data for an inclusive, in-plane and out-of-plane correlation function (filled circles) 
respectively. The harmonic correlations in each figure (dashed-dot curve) are calculated 
analytically from Eqn.~\ref{eq4} using $v_2$ from the fit in panel (a) along with 
$(v_2^A)^{out}$ and $(v_2^A)^{in}$ from Eqs.~\ref{eq11} and~\ref{eq12}. The ZYAM condition is used 
to fix the constant $a_0$ (in each case) as described above. Then the extracted points for 
jet-pair distributions are determined by difference via Eq.~\ref{eq7}. Comparison to the 
input jet-pair distribution (solid curves) shows an excellent agreement, thereby giving 
confidence in the decomposition procedure. It is noteworthy that a similar analysis performed 
for the case in which di-jets were made to correlate with the reaction plane, gave equally 
good results.

	As indicated earlier, this method of decomposition makes no assumption about the 
functional form of the jet function. From the shape of the jet functions observed in p+p and 
d+Au reactions, one is led to expect a double Gaussian jet function peaking at 0 and 180 deg. 
However, if the reaction creates a strongly interacting medium, then that medium could distort 
the jets, especially on the away-side. Such distortions have been recently attributed to 
to the influence of collective flow~\cite{Armesto:2004vz} and the generation of shock waves 
around partons propagating through the medium~\cite{Casalderrey-Solana:2004qm}. 
We have performed several simulations in which strongly distorted away-side jets
were introduced. In each case the decomposition method retrieves 
the input jet function in detail, confirming that the decomposition procedure is robust 
even for unusual di-jet distributions.

{\bf Extinction of Harmonic Correlations:}
Figure \ref{fig3}  shows pair correlations with strong harmonic 
components C$_{H}^{AB}(\Delta \phi )$ and relatively weak jet components 
J($\Delta \phi )$. Only after subtraction of C$_{H}^{AB}(\Delta \phi)$ is 
the true shape revealed for J($\Delta \phi )$. It would be desirable to 
select a data set that contained only those correlations due to the jets; 
Fig.~\ref{fig2} and Eqs.~\ref{eq5} and \ref{eq11} point the way. As discussed 
above for Fig.~\ref{fig2}, the out-of-plane constraint on particle A changes 
the phase of the harmonic correlation to favor pairs at $\Delta \phi $ = 90$^{o}$. 
By using Eq.~\ref{eq11} one can select the particular cut angle $\Delta\phi _{c}=\phi _{xt}$ 
such that (v$_{2}^{A})^{out}$ = 0. For this case the value of v$_{2}$ is 
driven to zero in Eq.~\ref{eq5}, and the harmonic correlations are extinguished.

\begin{figure}
 \includegraphics[width=1.0\linewidth]{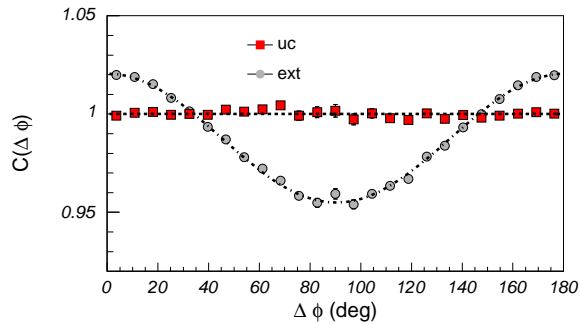}  
 \caption{\label{fig5}
 Simulated correlation function for unconstrained particles (filled circles) and for 
 a trigger particle constrained within the cut angle $\phi _{c} = \phi _{xt}$ perpendicular 
 to the reaction plane (filled squares), see text. The results are for a pure harmonic 
 simulation with $v_2 \sim 0.16$.
}
 \end{figure}

In Fig.~\ref{fig5} we demonstrate the extinguishing of harmonic correlations. 
The filled circles show the inclusive $\Delta \phi$ distribution obtained for a simulation
with an input $v_2 \sim 0.16$. The filled squares show the out-of-plane $\Delta \phi$ 
distribution after the cut angle $\phi _{c}$, is set to the extinction value, 
ie. $\phi _{c} = \phi _{xt}$. The latter distribution is flat and demonstrates 
that this technique gives a good method to determine the jet correlations 
directly from a data set, without the blurring effect from harmonic correlations 
mediated by the reaction plane. The latter technique provides clear advantages 
for future jet studies at RHIC. 

{\bf Jet pair fraction and conditional yield:}
The sum of the jet-pair function J over $\Delta \phi $ can be related to the average fraction of jet-correlated
particle pairs per event, hereafter termed jet pair fraction JPF, as follows,
\begin{equation}
JPF = \Sigma a_0 J (\Delta \phi ) / \Sigma C (\Delta \phi ). 
\label{eq16}
\end{equation}
The hatched area in \fig{fig1} illustrates this sum. This value of JPF represents 
the fraction of all mid-rapidity AB pairs that are correlated by jets i.e. 
those correlated by effects over and above those mediated by the reaction 
plane.

One may ask what is the event-averaged number of jet-associated particles
per trigger particle. This quantity is commonly referred to in jet-analyses
as per-trigger or conditional yield. The conditional yield can be calculated 
from the jet pair fraction as follows.

For a given data set, one records the detected number of AB pairs per event 
n$^{AB}$, the detected number of A particles per event n$^{A}$, and the 
detected number of B particles per event n$^{B }$. The ratio $R_p$ of the pair rates
to the product of the single particle rates 
\begin{equation}
 R_p = n^{AB}/(n^{A} \times n^{B}) 
\label{eq17a}
\end{equation}
carries information about pair-production over and above the combinatorial background.
These detected rates should be corrected 
for efficiencies to get the true rates for particle-pairs and singles at all 
$\phi $ but with the relevant $\eta $ acceptance.
Assuming that the pair-efficiency approximately factorizes into the 
single particle efficiencies, we can denote the true 
values per event by; 
\begin{eqnarray}
n^{A} = (n^{A}_{t})(Ef^{A}) 
\label{eq17}
\\
n^{B} = (n^{B}_{t})_{ }(Ef^{B})
\label{eq18}
\\
n^{AB} \approx (n^{AB}_{t})(Ef^{A})(Ef^{B}) 
\label{eq19}
\end{eqnarray}

i.e. $n_t^{AB}/n_t^An_t^B = n^{AB}/n^An^B$.

Now one writes for the true efficiency corrected conditional yield TCY
\begin{eqnarray}
TCY = JPF (n_t^{AB}/(n_t^{A} \times n_t^{B}))\times  n^{B}_{t} \\
 = JPF \times R_p \times n_t^B 
\label{eq20}
\end{eqnarray}
From the procedure outlined above one gets JPF, and one has $R_p$ from the 
detected values n$^{AB }$, n$^{A}$ and n$^{B}$ for each data set.
From the literature (see for example Refs.~\cite{Adcox:2003nr,Adler:2003cb}) 
one has previously published \pT spectra for many particles corrected for azimuthal 
efficiency. These can be integrated over the \pT intervals for 
particle B to get n$_{t}^{B }$.

	The evaluation of systematic errors in this procedure will center on the determination 
of the jet pair fraction (shaded area in Fig.~\ref{fig1} for example). This is because the 
jet-pair distribution could have a finite (albeit small) intensity at $\Delta\phi_{min}$, in 
certain situations. If this is indeed the case, the extracted jet shape is essentially 
unaffected, but the yields extracted under the ZYAM assumption would constitute lower 
limit yields. On the other hand, one can adopt an appropriate functional 
form for the shape of the jet function to make an estimate of the possible jet yield 
at $\Delta\phi_{min}$, and hence, assign a systematic error via iteration.

%

	In the discussion of jet yields we used \eq{eq19}
\[
n^{AB} = n_{t}^{AB}(Ef^{A}) (Ef^{B}). 
\]
In fact, we note that AB pairs that are correlated in $\Delta \phi$ may 
well have a somewhat modified efficiency, designated below by M$_{i}$.
\begin{equation}
n^{AB} = n_{t}^{AB}Ef^{A} Ef^{B} (M_{i})
\label{eq22}
\end{equation}
After determination of the harmonic amplitude one may get the M$_{H}$ factor 
by Monte Carlo simulation techniques. Similarly one can get modification 
factors M$_{nj}$ and M$_{aj}$ for the near-side jets and the away-side jets 
respectively.

Correlations between the multiplicities n$^{A}$ and n$^{B}$ make the mean number of combinatoric
pairs n$^{AB}_{comb}$ greater than the product 
n$^{A}$ n$^{B}$. One can write 
\begin{equation}
n^{AB}_{comb} / n^{A} n^{B} \xi = 1, \quad \xi \neq 1. 
\label{eq23}
\end{equation}
However the quantity $\xi$ cancels out in the product 
(JPF)(n$^{AB})$ in the present analysis.

{Summary:}
A Monte Carlo reaction code has been used to generate simulated data sets 
for hadron pair correlations tuned to the general behavior from Au + Au 
reactions at RHIC. These data sets contain pair correlations from both 
elliptic flow and from jets. A method is presented for the analysis of such 
data to retrieve the jet pair fractions, conditional yields, and widths 
for both near and away side jets. A novel technique is also presented for 
direct quenching of harmonic correlations to reveal jet correlations.


%
%

\bibliography{Methods}

\end{document}